\documentclass[singlecolumn,prd,amssymb,amsmath,preprintnumbers,floatfix,aps,nofootinbib]{revtex4-1}
\pdfoutput=1
\usepackage{bm,color}
\usepackage{slashed} 
\usepackage{graphicx}
\usepackage{soul} 
\usepackage{multirow}
\usepackage{tablefootnote}
\usepackage{comment}
\usepackage{epstopdf} 
\usepackage{mathtools}
\usepackage{array}
\usepackage{booktabs}
\usepackage{hhline}

\usepackage[colorlinks=true
,urlcolor=blue
,anchorcolor=blue
,citecolor=blue
,filecolor=blue
,linkcolor=blue
,menucolor=blue
,linktocpage=true
,pdfproducer=medialab
,pdfa=true
]{hyperref}

\newcommand{\beq}{\begin{equation}}
\newcommand{\eeq}{\end{equation}}
\newcommand{\bea}{\begin{eqnarray}}
\newcommand{\eea}{\end{eqnarray}}
\newcommand{\ba}{\begin{array}}
\newcommand{\ea}{\end{array}}

\def\m1{M_1}
\def\m2{M_2}
\def\m3{M_3}

\def\ch10{\tilde \chi^0_1}

\def\gev{\,{\rm GeV}}
\def\mev{\,{\rm MeV}}

\def\to{\rightarrow}

\def\cm{\,{\rm cm}}
\def\mm{\,{\rm mm}}
\def\km{\,{\rm km}}

\def\gram{\,{\rm gram}}

\newcommand{\lsim}{\mathrel{\mathop{\kern 0pt \rlap
  {\raise.2ex\hbox{$<$}}}
  \lower.9ex\hbox{\kern-.190em $\sim$}}}
\newcommand{\gsim}{\mathrel{\mathop{\kern 0pt \rlap
  {\raise.2ex\hbox{$>$}}}
  \lower.9ex\hbox{\kern-.190em $\sim$}}}

\def\fbi{\,{\rm fb}^{-1}}

\begin{document}
\title{\boldmath \bf \Large 
Millicharged Particles in Liquid Argon Neutrino Experiments
}
%Mapping the Higgs total width out from strong phases of the SM} 
\author{\bf
    Roni Harnik$^1$, Zhen Liu$^{2}$, and Ornella Palamara$^{1}$} 

\affiliation{\, \\
$^1$%Theoretical Physics Department, 
Fermi National Accelerator Laboratory, Batavia, IL 60510, USA\\
$^2$ Maryland Center for Fundamental Physics, Department of Physics, University of Maryland, College Park, MD 20742-4111 USA\\
%$^3$ INFN? Yale?
}

\date{\normalsize  \today}

\begin{abstract}
We investigate the potential of Liquid Argon (LAr) neutrino detectors to search for millicharged particles, a well-motivated extension of the standard model. Detectors located downstream of an intense proton beam that is striking a target may be exposed to a large flux of millicharged particles.  Millicharged particles interact primarily through low momentum exchange producing electron recoil events near detector threshold. Recently, sub-MeV detection capabilities were demonstrated by the Fermilab ArgoNeuT detector, a small LAr detector which was exposed to the NuMI neutrino beam. Despite high background rates and its small size, we show that ArgoNeuT is capable of probing unexplored parameter space with its existing dataset. In particular, we show that the excellent spatial resolution in LAr detectors allows rejecting backgrounds by requiring two soft hits that are aligned with the upstream target. We further discuss the prospects of these types of searches in future larger LAr neutrino detectors such as the DUNE near detector.
\end{abstract}

\preprint{
%\begin{flushright}
FERMILAB-PUB-19-060-ND-T
%\end{flushright}
}

\maketitle
%%%%%%%%%%%%%%%%%%%%%%%%%%%%%%%%%%%%%%%%%%%%%%%%%

%\flushbottom 

\section{Introduction}

A vital component in the quest for new particles beyond the Standard Model (SM) is the search for new light states, at or below the GeV scale, that are very weakly coupled. A particularly simple possibility that realizes this is the introduction of new particles that carry a small electric charge, so-called millicharged particles (mCPs). In their simplest form, they may be introduced as just that, new particles that violate the quantization of charge seen in the SM. Millicharged particles can also arise more elegantly in the low energy limit of a theory in which a new dark photon kinetically mixes with the visible one~\cite{Holdom:1985ag}. Millicharged particles could make up part of the dark matter in the Universe~\cite{Brahm:1989jh, Boehm:2003hm, Foot:2004pa, Pospelov:2007mp, Feng:2009mn, Kaplan:2009de, Cline:2012is, Tulin:2012wi, Agrawal:2016quu} and this possibility has recently attracted attention in the context of the EDGES 21 cm anomaly~\cite{Barkana:2018lgd, Berlin:2018sjs, Barkana:2018qrx}.

In this work, we propose to search for mCPs in liquid argon (LAr) detectors in neutrino beams. 
We will show that ArgoNeuT~\cite{Anderson:2012vc}, one of the first and smallest such detectors, can already probe new regions of the mCP parameter space with existing data.
Millicharged particles can be produced at any intense fixed target setup via the decay of mesons or through bremsstrahlung. In particular, they will be produced in the target that is struck by a proton beam to produce neutrinos. Due to their small charge, mCPs will, for the most part, travel in an approximately straight path through magnetic fields and shieldings, traversing the neutrino near detectors.

\begin{figure}[b]
    \begin{center}
        \includegraphics[scale=0.4,clip]{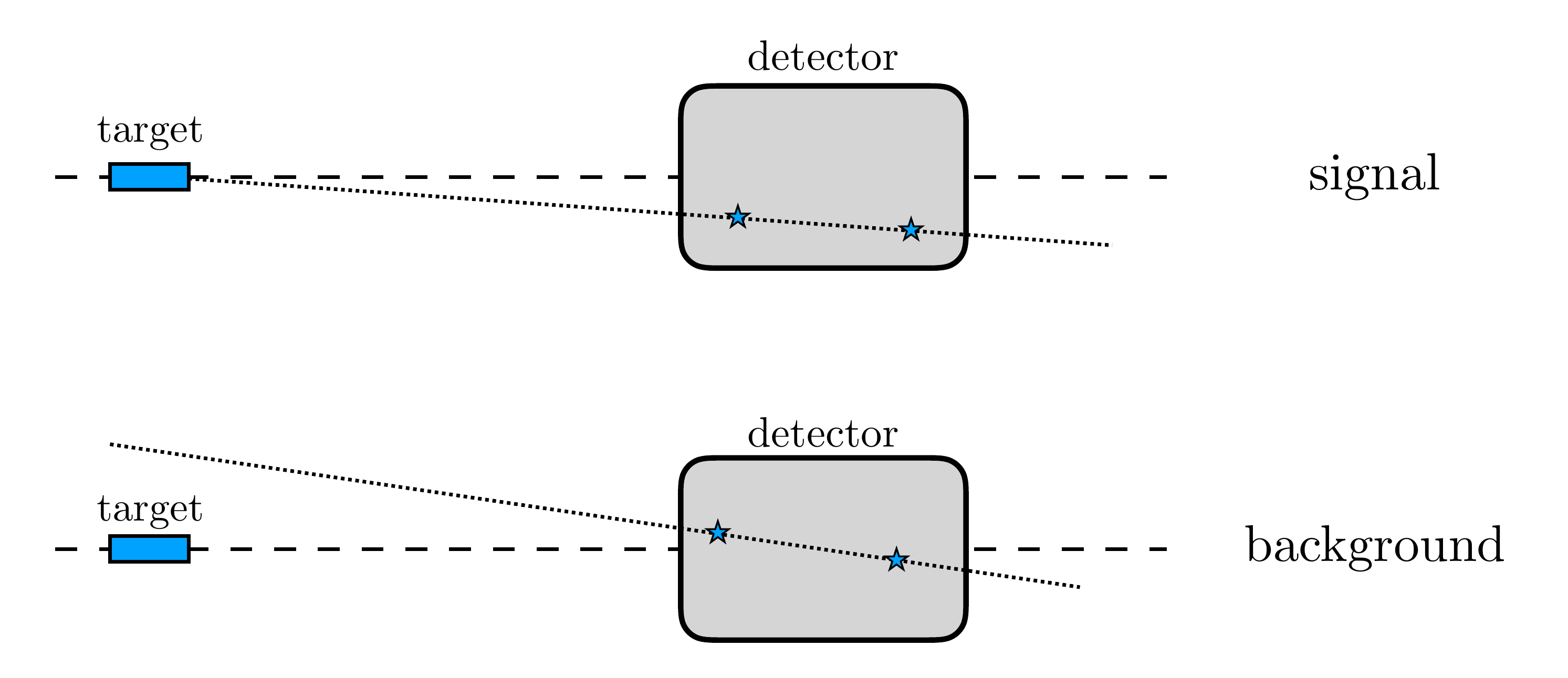}
        \caption{In a signal double-hit event, the line that is defined by the two hits will point to the target~(top), whereas in a background double-hit event, it generically will not~(bottom). Searching for double-hit events can thus lead to an enhanced signal to background ratio. 
        }
        \label{fig:cartoon}
    \end{center}
\end{figure}

LAr detectors are well suited to search for these particles. At the microphysical level, high energy millicharged particles interact with matter similarly to their charged counterparts through soft ionizing collisions, though with reduced rates. It has recently been shown that LAr detectors can resolve the individual collisions down to a threshold of around MeV or less~\cite{Acciarri:2018myr}. The mCP signal thus consists of one or more soft hits within the detector volume. 
In the case of two or more hits, we will show that signal events will be aligned with the target in which the mCPs were produced as shown in Figure~\ref{fig:cartoon}. 
In contrast, background double hit events will be uniformly distributed in the detector volume and will only rarely align with the target. A central result of this work is that searching for two or more hits that are in line with the production target can be used to reduce the backgrounds and improve the reach effectively. 

Laboratory-based limits on mCPs have been placed by the milliQ experiment at SLAC~\cite{Prinz:1998ua}. More recently more search strategies have been proposed, both in a dedicated experiment at LHC (milliQan)~\cite{Ball:2016zrp} and other fixed-target setups~\cite{Berlin:2018bsc}. During the preparation of this work, Ref.~\cite{Magill:2018tbb} has also suggested a search for mCPs in neutrino detectors, including miniBooNE, microBooNE, and SBND using the booster 8 GeV beam line and with the future DUNE LAr Near Detector (DUNE ND)\footnote{
Ref.~\cite{Magill:2018tbb} only considered single-hit events. Our work generally agrees with the signal rates in this study, though we will take a more conservative approach to backgrounds in LAr and assume uncertainties are systematic in nature. To this end, the use of the multi-hit signal for suppressing backgrounds will allow for a nearly background-free search. An additional difference is that here we consider the use of the existing 120 GeV NuMI beam line with ArgoNeuT, which will cover a sizable new parameter regime already.}.  

The paper is structured as follows. In section~\ref{sec:prod} we discuss the production of mCPs in the NuMI beamline (120 GeV) which primarily proceeds via meson decay and bremsstrahlung for high masses.  We will also consider the small matter effects on mCPs en route to the detector. In Section~\ref{sec:detect} we describe the interaction of mCPs in liquid Argon and consider the signal rate for single- and double-hit events. In Section~\ref{sec:ArgoNeuT} we discuss the background rates in ArgoNeuT and estimate the reach of a dedicated analysis with existing data. In Section~\ref{sec:DUNE} we discuss some of the opportunities and challenges of searching for mCPs in larger detectors that are also exposed to more intense beams and in Section~\ref{sec:outlook} we conclude. 

%The strategy for millicharge detection is shaped by their behavior as they pass through matter.
%Like charged particles, most collisions of mCPs are soft ionizing collisions. For charged particles, these collisions are very frequent and quantities such as $dE/dx$, which effectively sum over the multiple soft collisions, are useful. 
%For mCPs, where these collisions are fewer and far between this description may have less utility.
%In particular, for detectors that have high spatial resolution such as those we consider here, the mean free path between individual interactions can be    
%A detector for millicharged particles would thus be most sensitive if it is able to detect soft energy depositions.

%As a result, the $dE/dx$ description which is of great utility for charged particles is not always the most useful treatment

\section{Millicharge Production and Acceptance in ArgoNeuT and DUNE ND}
\label{sec:prod}

Accelerator-based neutrino beams are sourced by a high-intensity proton beam striking a fixed target and may thus also be an excellent source mCPs. In this work, we focus on the NuMI beam line and the future LBNF, both of which include 120 GeV protons striking a target. 
The center of mass energy of the proton-proton collision is hence,
\beq
s\sim(15~\gev)^2.
\eeq
Using a high energy proton beam (as compared, say to the 8 GeV BNB) allows for mCP production up to higher energies and will also produce a more focused mCP, enhancing the geometric acceptance.

Of particular interest in this work is the existing dataset collected by the ArgoNeuT experiment~\cite{Acciarri:2018myr}.
%one existing dataset and experiment ArgoNeuT and one dominant future neutrino experiment DUNE, both of which run at a beam energy of 120 GeV. 
The ArgoNeuT detector is a small LAr TPC of dimensions
\begin{equation}
%H\times W \times L = 0.47 \times 0.4 \times 0.9\ \mathrm{meter}^3
\mbox{ArgoNeuT:} \qquad \Delta y\times \Delta x \times \Delta z = 0.40 \times 0.47 \times 0.90\ \mathrm{meter}^3
\eeq
which was located 975 meters downstream from the NuMI target and was approximately on axis. Here $\hat y$ is the vertical derection in the lab and $\hat z$ is horizontal and roughly in the beam direction.
The existing ArgoNeuT dataset consists of about $10^{20}$ protons on target (POT).
We will consider larger detectors in Section~\ref{sec:DUNE}. In particular, we will estimate the sensitivity of the future DUNE ND. 
We assume DUNE ND to be of dimensions~\cite{DUNEND}
\beq
\mbox{DUNE ND:} \qquad \Delta y\times \Delta x \times \Delta z =3\times 4\times 5~\mathrm{meter}^3 
\eeq
and placed 574 meters from the target with a total exposure of $3\times10^{22}$ POT over its full run. 
%We tabulate the relevant parameters in Table.~\ref{tab:geometry}.
In this section, we compute the flux of mCPs that may have passed through ArgoNeuT and may pass through DUNE ND for a given mCP charge and mass. 

The mCPs can be produced directly from photon mediated processes at low energy. These include millicharged particles from decays of scalar mesons such as $\pi^0$, $\eta$, $\eta^\prime$, vector mesons such as $\rho$, $\omega$, $\phi$, $J/\psi$ and direct Drell-Yan (DY) productions.
We have generated mCP production using {\tt Pythia8}~\cite{Sjostrand:2006za,Sjostrand:2007gs} and  validated the DY sample using {\tt Madgraph5}~\cite{Alwall:2011uj}. On average, about three neutral pions are produced per Proton-On-Target (POT) and a few orders of magnitude less other mesons.\footnote{The number of mesons production per POT is defined as the following $\sigma_{pp\to {\rm meson}+X}/\sigma_{\rm total}$. 
Here $\sigma_{pp\to {\rm meson}+X}$ is the inclusive production rate of mesons in proton-proton collisions. The cross section in the denominator $\sigma_{\rm total}$ is the soft QCD total cross section which is around 40 milli-barn which dominates over the hard and non-diffractive scattering.}
Although for every hard QCD collision one produces many SM hadrons, e.g., 9.3 $\pi^0$ per collision, the cross section is dominated by the soft QCD process~\cite{Rasmussen:2018dgo}.
The $\pi^0$, $\eta$, $\eta^\prime$ mesons decay via three-body channels into a photon plus a pair of mCPs. The $J/\psi$ decays into a pair of mCPs directly via its mixing with the photon. The DY process produces pairs of mCPs directly as well. 
The corresponding meson production rates and branching fraction into mCP pairs are shown in Table.~\ref{tab:acceptance} in the massless mCP limit.
\begin{table}[tb]
    \centering
    \caption{Meson production rates at a 120 GeV proton fixed target facility and their branching fractions to pairs of light mCPs, and the number of mCP pairs from DY production. The phase space suppression factors for branching fraction calculations are further implemented in our calculation for heavier mCPs. The geometric acceptance $A_\mathrm{geo}$ of ArgoNeuT and DUNE ND for each production mode are shown for two mCP mass benchmarks
    %\ZL{The lower panel is the case for DUNE, we shall decide whether to merge it with ArgoNeuT or to present it in a later section.}
    . %{\bf RH: make sure i interpreted the factor of 2 correctly.}
    }
    \bgroup
         \def\arraystretch{1.25}%  1 is the default, change whatever you need
      \begin{tabular}{rcccccccc}
      \hline \hline
       \qquad & $\pi^0$ & $\eta$ & $\eta^\prime$ & $\rho$ & $\omega$ & $\phi$ & $J/\psi$ & DY \\
      \hline %\\[-8pt]
\#/POT & $2.9$ & $3.2\times 10^{-1}$ & $3.4\times 10^{-2}$ & $3.7\times 10^{-1}$ & $3.7\times 10^{-1}$ & $1.1\times 10^{-2}$ & $5.4\times 10^{-7}$ & $4.7\times 10^{-10}\epsilon^2$ \\
 $2\times \mathrm{Br}_{X\to\chi\bar\chi}(\%)$ &   $2.3 \epsilon^2$   & $1.4\epsilon^2$  & $0.04\epsilon^2$ & $0.009\epsilon^2$ & $0.018\epsilon^2$ & $0.058\epsilon^2$ & $12\epsilon^2$ & ---   \\
$A^{\rm ArgoNeuT}_\mathrm{geo}(m_\chi$=20 MeV) & $3.1\times 10^{-5}$ & $2.1\times 10^{-5}$  & $1.6\times 10^{-5}$ & $1.9\times 10^{-5}$ & $2.0\times 10^{-5}$ & $9.1\times 10^{-6}$  &  $5.0\times 10^{-6}$ & $3.2\times 10^{-6}$  \\
 $A^{\rm ArgoNeuT}_\mathrm{geo}(m_\chi$=200 MeV) & --- & $5.4\times 10^{-5}$  & $3.4\times 10^{-5}$ & $2.3\times 10^{-5}$ & $2.2\times 10^{-5}$ & $1.1\times 10^{-5}$  &  $4.6\times 10^{-6}$ & $3.1\times 10^{-6}$  \\%[2pt]
%       \hline\hline
%        \qquad & $\pi^0$ & $\eta$ & $\eta^\prime$ & $\rho$ & $\omega$ & $\phi$ & $J/\psi$ & DY \\
%       \hline \\[-8pt]
% \#/POT & $4.5$ & $5.1\times 10^{-1}$ & $5.4\times 10^{-2}$ & $5.8\times 10^{-2}$ & $5.9\times 10^{-2}$ & $1.7\times 10^{-2}$ & $8.7\times 10^{-7}$ & $7.4\times 10^{-10}\epsilon^2$ \\
%  $2\times \mathrm{Br}_{X\to\chi\bar\chi}(\%)$ &   $2.3 \epsilon^2$   & $1.4\epsilon^2$  & $0.04\epsilon^2$ & $0.009\epsilon^2$ & $0.018\epsilon^2$ & $0.058\epsilon^2$ & $12\epsilon^2$ & ---   \\
$A^{\rm DUNE}_\mathrm{geo}(m_\chi$=20 MeV) & $6.5\times 10^{-3}$ & $5.2\times 10^{-3}$  & $3.7\times 10^{-3}$ & $4.2\times 10^{-3}$ & $4.2\times 10^{-3}$ & $2.2\times 10^{-3}$  &  $1.1\times 10^{-3}$ & $6.7\times 10^{-4}$  \\
 $A^{\rm DUNE}_\mathrm{geo}(m_\chi$=200 MeV) & --- & $1.0\times 10^{-2}$  & $5.3\times 10^{-3}$ & $5.0\times 10^{-3}$ & $2.3\times 10^{-3}$ & $1.1\times 10^{-3}$  &  $6.6\times 10^{-4}$ & $3.1\times 10^{-6}$  \\[2pt]
      \hline\hline
%      \multicolumn{1}{|c|}{\multirow{2}[0]{*}{DUNE near}} &$m_\chi$=20 MeV & $2.8*10^{-3}$ & $2.5*10^{-3}$ & $2.1*10^{-3}$ & $1.1*10^{-3}$ & $6.7*10^{-4}$ \\
%  \cline{2-7}    \multicolumn{1}{|c|}{} & $m_\chi$=200 MeV & --- & $4.3*10^{-3}$ & $2.9*10^{-3}$ & $1.1*10^{-3}$ & $6.7*10^{-4}$ \\
%      \hline
      \end{tabular}%
      \egroup
    \label{tab:acceptance}%
\end{table}%
In the table, we also show $A_\mathrm{geo}$, the fraction of mCPs produced by the various sources that go through the ArgoNeuT detector and DUNE ND. The small ArgoNeuT detector covers a solid angle of $\sim 2\times 10^{-8}$ of the full sphere as viewed from the NuMI target. However, geometric acceptance is much higher due to the significant boost of produced mCPs. The DUNE ND is larger in angular coverage of around $4\times 10^{-6}$ of the full sphere. As the mass of the mCP is taken to be higher, the various meson decays into $\chi\bar\chi$ pairs become kinematically inaccessible, and the overall flux of mCPs at ArgoNeuT and DUNE ND is reduced. This is shown in of Figure~\ref{fig:prod} where we plot the number of mCPs traversing ArgoNeuT (left panel) and DUNE ND (right panel) over its full run as a function of the mCP mass $m_{\chi}$ \footnote{Heavy ($\geq$ GeV) mCPs are produced through high $x$-values of the parton distribution functions, whose uncertainty will impact the rate estimation. These can be included in an experimental study of the ArgoNeuT dataset.}.
Because DUNE ND has around 200 times more angular acceptance than ArgoNeuT, the flux of mCPs from different sources increases by a different amount due to detailed kinematics, which changing the relative strength of mCPs slightly from different sources.
\begin{figure}[t]
    \begin{center}
        \includegraphics[scale=0.55,clip]{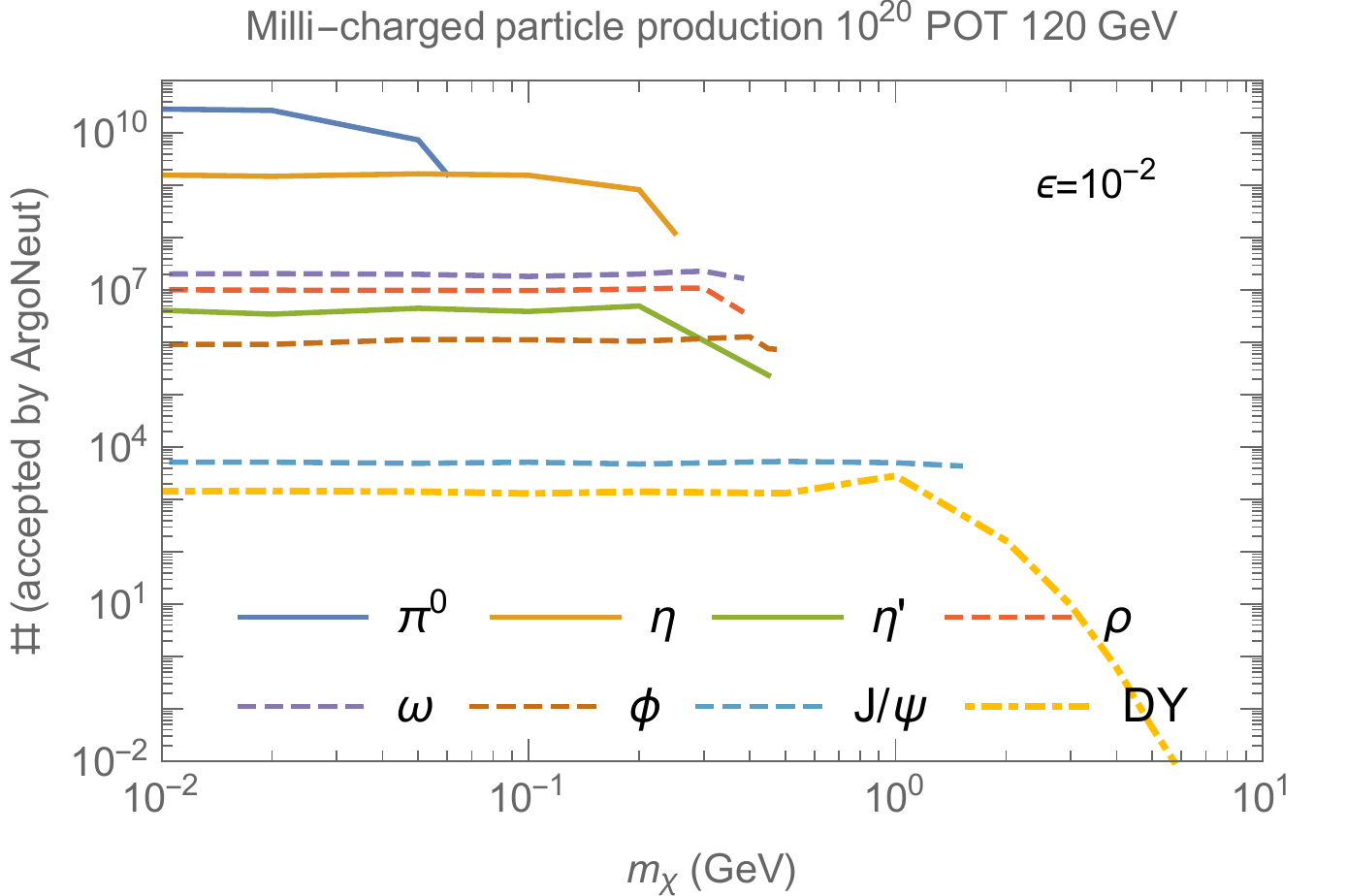} 
        \includegraphics[scale=0.55,clip]{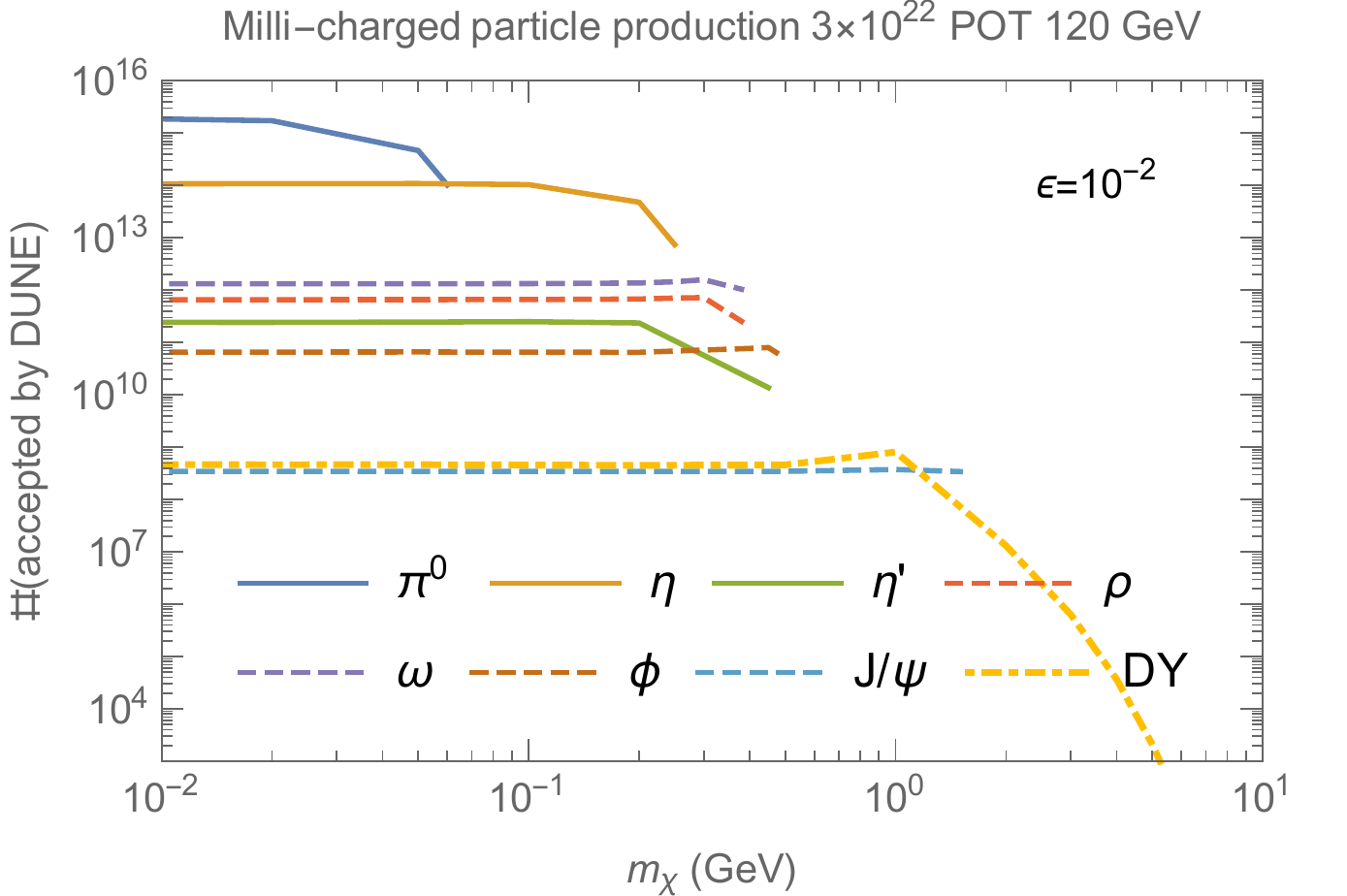}
        \caption{The number of millicharged particles produced through various sources within ArgoNeuT's~(left) and DUNE ND's~(right) geometrical acceptance for $\epsilon=10^{-2}$, integrated over all mCP energies. We assume $10^{20}$ and $3\times 10^{22}$ POT for ArgoNeuT and DUNE ND respectively. The mCP production from scalar mesons $\pi$, $\eta$, $\eta^\prime$ are shown in solid lines from top-down. The mCP production from vector mesons $\rho$, $\omega$, $\phi$, $J/\psi$ are shown in dashed lines from top-down, and the Drell-Yan production is shown in dot-dashed line. The production rate scales as $\epsilon^2$.
        %{\bf RH:  remove the /yr from y-axis label.}
        } 
        \label{fig:prod}
    \end{center} 
\end{figure}
Interestingly, for the DY process, a mild increase in the detector geometrical acceptance can be found for mCP masses at around 1 GeV. This increase is because at high masses mCP's are produced near threshold, and thus have limited transverse momentum. This increase also holds for mCPs at the threshold of the parent particle (half the parent meson mass). However, the phase-space suppression at threshold is stronger such that this increase in acceptance is not visible in Figure~\ref{fig:prod}.

To assess the effect of soft scatterings on the trajectory of the mCP, it will be useful to study their typical energies. In Figure~\ref{fig:energydistribution} we show the energy distribution of mCPs that are accepted by ArgoNeuT and are produced through decays of neutral pions, $\omega$'s and through the DY process. In each case, we choose mCP masses in a range that the corresponding production source accounts for a high fraction of the overall production. For lower mCP masses and a given production mode, the energy distributions are very similar to each other as the mass is a tiny correction to the kinematics, as is shown by the blue curves in each panel.
\begin{figure}[t]
    \begin{center}
        \includegraphics[scale=0.55,clip]{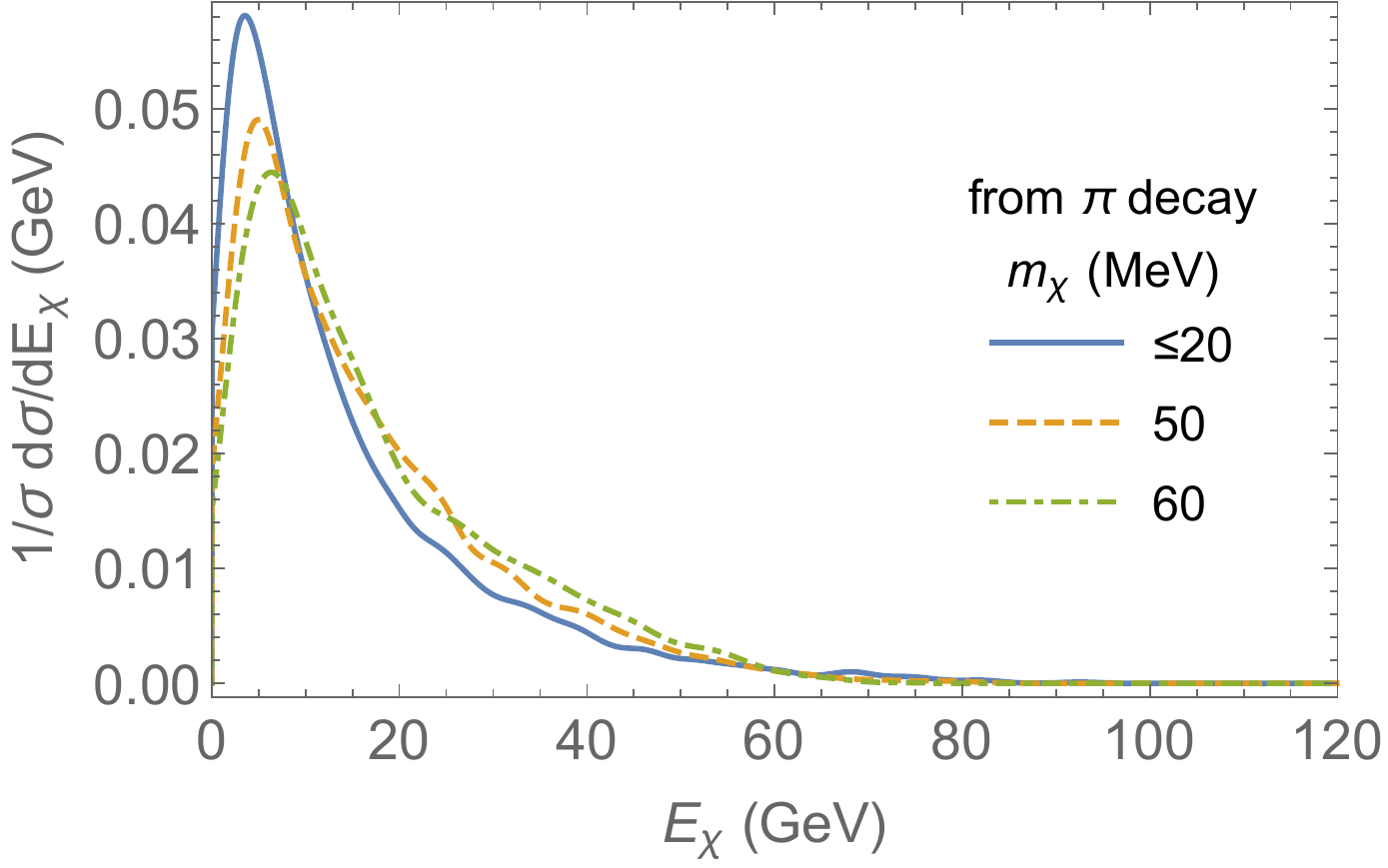}
        \includegraphics[scale=0.55,clip]{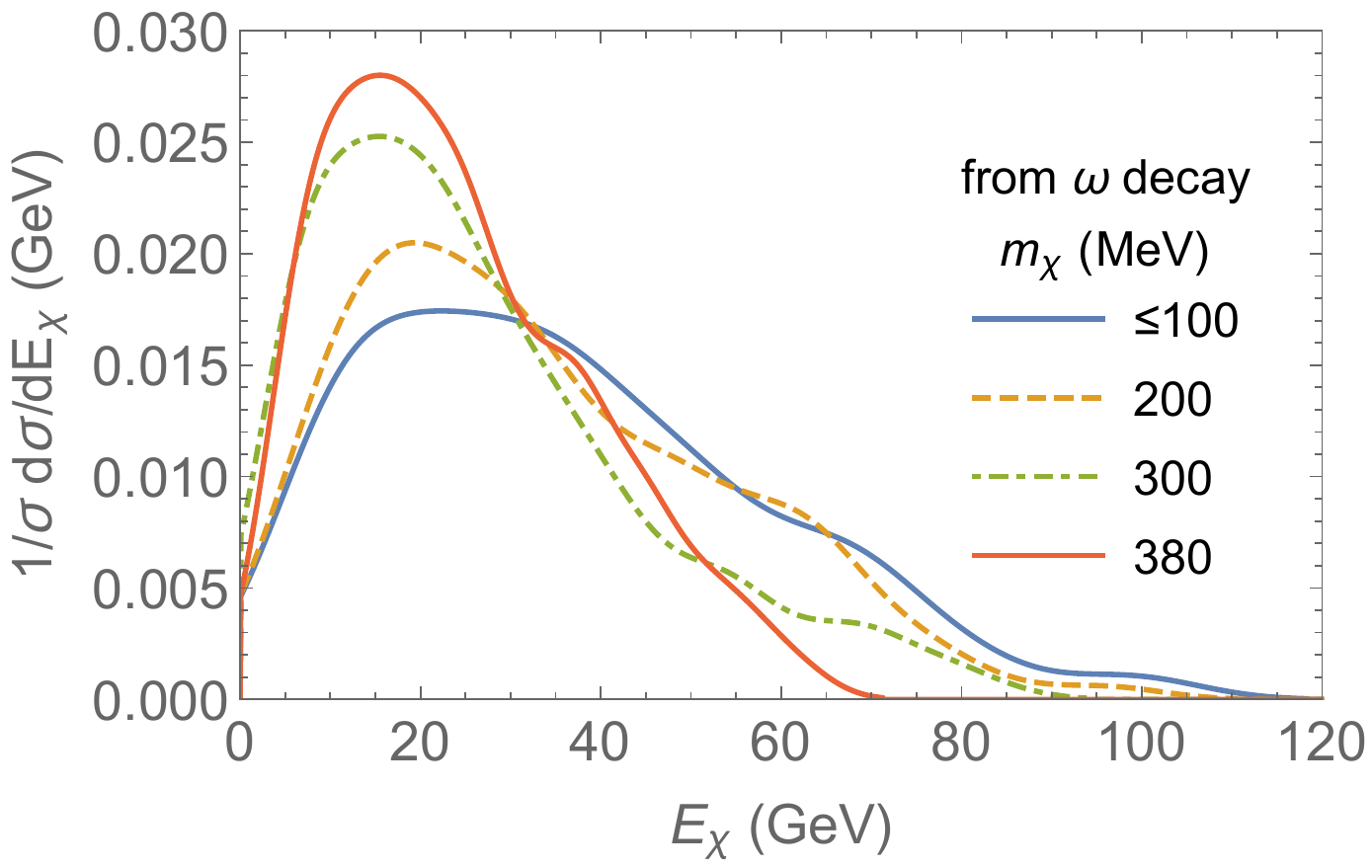}
        \includegraphics[scale=0.55,clip]{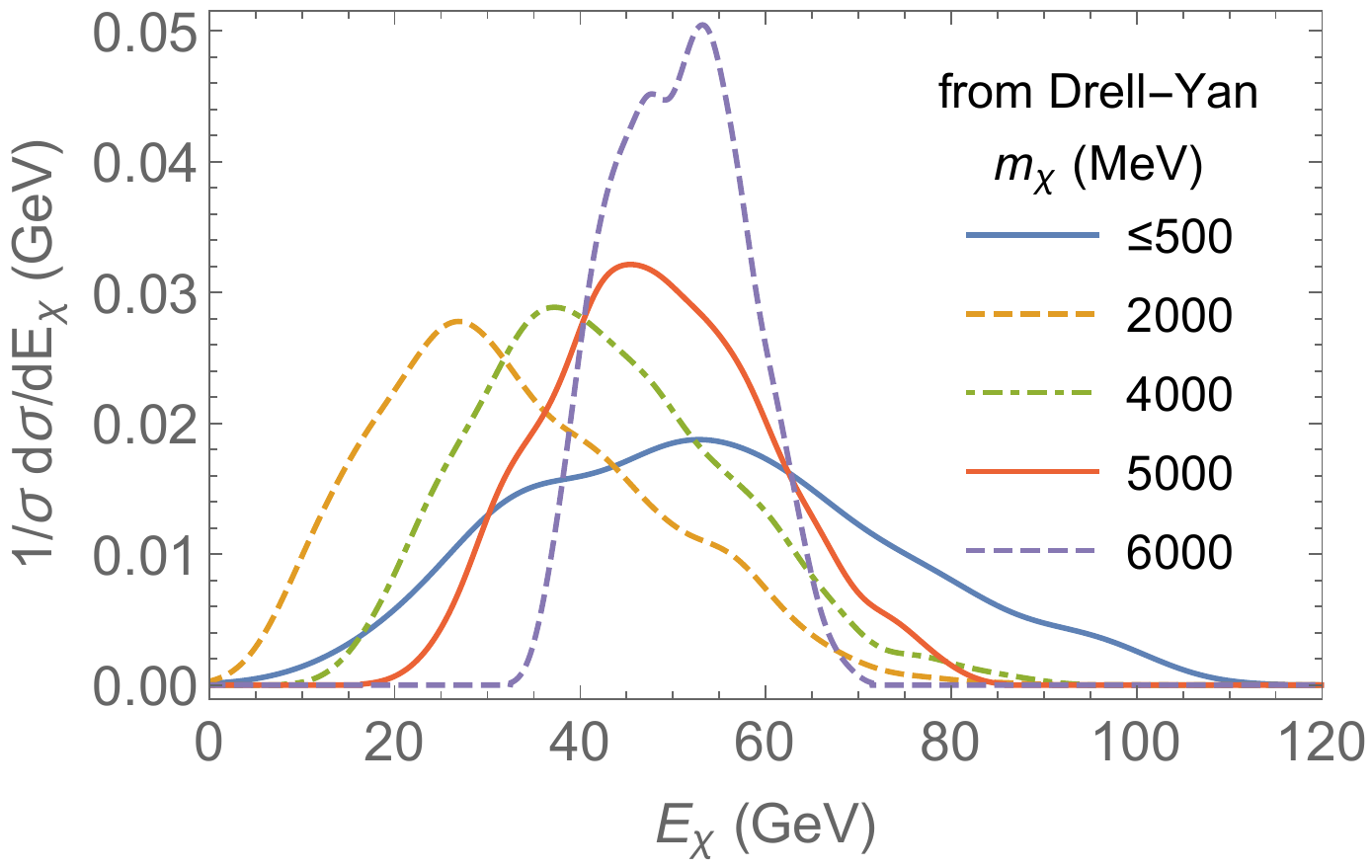}
        \caption{The energy distribution of millicharged particles that are accepted by ArgoNeuT from various sources and for different masses. In each plot, mCP masses are chosen so that the respective production mode is dominant. 
        %\footnote{}
        }
        \label{fig:energydistribution}
    \end{center}
\end{figure}
This figure shows that in the region of interest the mCPs that go through on-axis near detectors will be highly relativistic, often possessing over 10~GeV of energy. 
%Note that the heavier the mother particle is, Furthermore, the lower the emittance it will have in the center of mass frame. This argument also applies to the DY process.
We can thus conclude that a high flux of relativistic mCPs is produced in neutrino beams motivating a search for them in near detectors.

\subsection{Matter Effects} \label{sec:angle}

Before discussing the detection of mCPs in LAr detectors, we briefly discuss the effect of the roughly 500 meters of matter which they traverse between the target and the detector. Matter can affect mCPs in two ways - causing energy loss and changing its direction~\cite{Tanabashi:2018oca}. Energy loss is dominated by the interaction of mCPs with electrons. Of these, the rare hard collisions dominate energy loss while soft collisions dominate the number of scatterings. As a result of the rare hard collision dominating energy loss and the collision probability decreasing as inverse squared of the recoil energy\footnote{for more details, see the next section.}, an estimate of the typical energy loss by a mCP in 500~meters of dirt can thus be estimated by solving for the minimal recoil energy at which the mean free path equals 500~meters. As we shall see in the next section, where mean free paths are calculated, this amounts to a typical energy loss of a few MeV.  This energy loss estimate agrees with the estimate from the well known Bethe-Block formula for $dE/dx$~\cite{Tanabashi:2018oca}\footnote{ Using the canonical Bethe Block formula gives a slightly higher energy loss, of order 10~MeV. The difference can be reconciled because Bethe-Block is the average energy loss per unit radiation length in an asymptotically long path through matter. For our case, in which we are interested in a finite path of order a kilometer, the ``typical energy loss'' is a more appropriate quantity. This effect amounts to a decrease in the estimated loss by up to an order of magnitude~\cite{Tanabashi:2018oca}.}.  
%The $dE/dX$ for a relativistic particle of charge $10^{-2} e$ is of order ${10^{-4} \rm MeV\times cm^2/g}$. For 500 meters of dirt with an average density of $2.6~{\rm g/cm^3}$, the typical energy loss is around $13~\mev$. This energy loss is far below the typical energy of 10s of GeV of the accepted mCPs as shown in Fig.~\ref{fig:energydistribution}.  
We thus conclude that matter causes an unobservable change in the energy spectrum in the region of interest.

The second effect, that of directional change, is dominated by scattering off of nuclei~\cite{Tanabashi:2018oca}. This too is a small effect, but requires attention since we will make use of the fact that mCPs arrive to the detector from the direction of the target. Since the estimate of the deflection angle requires some of the formulae in the next section, we will defer the details to Appendix~\ref{sec:deflection}. The result, however, is that a typical mCP is deflected by traversing a length $L_\mathrm{dirt}$ of dirt by an angle
\beq
\label{eq:matter}
\Delta \theta_\chi \sim 2\times 10^{-3} 
\left(      \frac{5\ \mathrm{GeV}}{E_\chi}  \right) \left(  \frac{\epsilon}{10^{-2}}  \right) \left( \frac{L_\mathrm{dirt}}{500\ \mathrm{meters}}\right)^{1/2}\,. 
\eeq

Given the fixed target phase space is highly collimated in the forward direction, the fraction of signals deflected into the detector would be smaller compared to the number of signals being deflected away from ArgoNeuT, which is very tiny. 
%Since we have estimated for the signal the typical energy loss is less than 0.1\%, the flux to be deflected in will be of the same order. 
Furthermore, the signal deflected into the detector will not be extrapolated back to be aligned with the target and would not be counted as signals in our setup. Consequently, we neglect the number of additional mCPs that are deflected and accepted by the ArgoNeuT experiment.

Having established that mCP retain their energy spectrum and direction in going through matter, we are now ready to consider their interactions in a LAr detector.

\section{Millicharge Interactions in Liquid Argon}
\label{sec:detect}

We have seen that millions to billions of mCPs may pass through neutrino near detectors during their exposure. In this section, we will consider strategies to detect them using electron recoils, including single and double hit events near the detector threshold of order MeV. We begin with calculations of the scattering rate and the mean free path of mCPs in LAr.

\subsection{Cross section and mean free path}

Given that millicharged particles have a low cross section and large mean free path, a scattering treatment for the detection and interaction with the material would be in general appropriate. The cross section is most significant for scattering off of electrons which will be discussed here. Though we will be interested in low recoil energies, of order MeV, the momentum exchange is sufficiently large compared to atomic scales, and we will approximate the electrons to be free.
The differential cross section for a millicharged particle to scatter off an electron is
\beq
\frac {d\sigma} {dE_r}= {\pi \alpha^2 \epsilon^2}\frac {2E_\chi^2 m_e + E_r^2 m_e-E_r \left(m_\chi^2+m_e(2 E_\chi+m_e)\right)} {E_r^2 (E_\chi^2-m_\chi^2)m_e^2},
\label{eq:dsigmader}
\eeq
where $E_r$ 
%\equiv E_e^f-m_e$ 
is the recoil energy defined in the lab frame assuming initial electron is stationary. $E_\chi$ and $m_\chi$ are the initial millicharged particle $\chi$ energy and mass, respectively. In the ultra-relativistic limit, equation~(\ref{eq:dsigmader}) reduces to
\beq
\left.\frac {d\sigma} {d E_r}\right\rvert_{E_\chi\gg m_\chi, m_e, E_r}\simeq \frac {2\pi\alpha^2 \epsilon^2} {E_r^2 m_e}.
\label{eq:dsigmader_rel}
\eeq
The differential cross section diverges when $E_r$ approaches zero as is well known for Coulomb scattering.
The total cross section after integrating  equation~(\ref{eq:dsigmader}) over the range of recoil energy $E_r$ of $[E_r^\mathrm{min},E_r^\mathrm{max}]$ is 
\beq
\sigma(E_r^\mathrm{min},E_r^\mathrm{max})=\pi \alpha^2 \epsilon^2 \frac {m_e(E_r^\mathrm{max}-E_r^\mathrm{min})(2 E_\chi^2 +E_r^\mathrm{max}E_r^\mathrm{min}) - E_r^\mathrm{max}E_r^\mathrm{min}\left(m_\chi^2 +m_e (2E_\chi+m_e)\right)\log\frac{E_r^\mathrm{max}}{E_r^\mathrm{min}}} {E_r^\mathrm{max}E_r^\mathrm{min} (E_\chi^2-m_\chi^2)m_e^2}\,.
\label{eq:sigma}
\eeq
%\beq
%\sigma(E_r^\mathrm{min},E_r^\mathrm{max})=
%\pi \alpha^2 \epsilon^2 
%\frac {(E_r^\mathrm{max}-E_r^\mathrm{min})(2 E_\chi^2 +E_r^\mathrm{max}E_r^\mathrm{min})m_e - E_r^\mathrm{max}E_r^\mathrm{min}left(m_\chi^2 +m_e (2E_\chi+m_e)\right)%\log\frac{E_r^\mathrm{max}}{E_r^\mathrm{min}}}{E_r^\mathrm{max}E_r^\mathrm{min} (E_\chi^2-m_\chi^2)m_e^2}\,.
%\label{eq:sigma}
%\eeq
When there is no upper cut placed on the recoil energy the upper limit  can be replaced by the maximal kinematically allowed recoil energy
\beq
E_r^\mathrm{max}=\frac {(E_\chi^2-m_\chi^2)m_e}{m_\chi^2+2E_\chi m_e + m_e^2}\,.
\label{eq:ermax}
\eeq
The total cross section scales as $E_r^\mathrm{min}$ and is thus dominated by soft collisions. %These most common interactions are 

Given a detector threshold for electron recoil, 
the scattering cross section in Eq.~(\ref{eq:sigma}) can then be conveniently converted to a mean free path,
\beq
\lambda(E_r^\mathrm{min})= \frac{1}{Z n_\mathrm{det} \sigma(E_r^\mathrm{min})},
\label{eq:meanfreepath}
\eeq
where $Z$ is the atomic number of the target element in the detector and $n_\mathrm{det}$ is its number density.
With the demonstrated capability of ArgoNeuT to detect sub-MeV scattering~\cite{Acciarri:2018myr} we will be interested in thresholds of that order.
For liquid argon 
%that is used at ArgoNeuT and many other neutrino experiments, 
$\rho=1.3954 {\rm g}/\cm^3$, $Z=18$, $m_a=39.948\gram/\mathrm{mol}$, the mean free path for an ultra-relativistic millicharged particle to scatter above threshold is
\beq
\lambda(E_r^\mathrm{min})\simeq  \left(\frac {10^{-2}} {\epsilon}\right)^2 \left(\frac {E_r^\mathrm{min}} {1~\mev}\right)\ 1\, {\km}.
\label{eq:lambda}
\eeq
Before proceeding to propose a search strategy we point out that soft scattering, those with electron recoils of order MeV, lead to a very small angular deflection of the mCP.  The deflection angle is of order the momentum transfer in the collision divided by the initial momentum of the mCP
\beq
\label{eq:scattering-angle}
\theta_\chi \sim \frac{\sqrt{2 m_e E_r}}{E_\chi}\,.
\eeq
The momentum transfer, $\sqrt{2 m_e E_r}$, is of order MeV, while the initial momentum of the typical mCP is of order one to tens of GeV, as shown in Fig.~\ref{fig:energydistribution}. 
The deflection angle of a beam mCP which deposits an observable hit in a LAr detector will be thus deflected by an angle that is at most $10^{-3}$. We conclude that after an observable hit the mCP will maintain its trajectory and that any further hits will be along a line pointing back to the target. As shown in Figure~\ref{fig:cartoon}, we will use this to identify signal events and suppress backgrounds.

\subsection{Interaction probabilities - one and two hits}\label{sec:probabilities}

The probability of an mCP to interact within a detector of size $L$ is $L/\lambda$ when the mean free path is larger than the detector, which is valid for small millicharges of order $10^{-2}$ or below. Since we would like the limit setting procedure to be smooth as we go to larger millicharges and physically large detectors, we will write a probability for interaction that is valid even when the mean-free-path is of order the detector size or smaller. To this end, we will consider a segmented detector with a spatial resolution of $\Delta L$ in the beam direction. The probability to scatter once above a threshold 
$E_{r}^\mathrm{min}$ within a detector segment of  size $\Delta L$ along the mCPs path is
\begin{equation}
p = \frac{\Delta L}{\lambda}\,
\end{equation} 
where $p$ implicitly depends on the threshold, the mCPs charge as in equation~(\ref{eq:lambda}), and depend more weakly on the mCP mass and energy via equation~(\ref{eq:sigma}).
The probability to interact $n$ times in a detector of length $L=N_L\Delta L$ is thus
\begin{equation}
P^{(n)}_{L}= \begin{pmatrix} n \cr N_L \end{pmatrix} p^n = \frac{N_L!}{n!(N_L-n)!}\left(\frac{\Delta L}{\lambda}\right)^n\,.
\end{equation}
It is easy to see that in the limit $\lambda\gg L$ this reduces to
\begin{equation}
P^{(n)}_{L}\sim \frac{1}{n!}\left(\frac{L}{\lambda}\right)^n\,,
\end{equation}
as expected. 
%
%Considering two hit event, which are expected to be in line with the distant target (see Figure~\ref{fig:cartoon}) will allow for a large reduction of background. 
Calculating event rates for one or two hits requires folding these interaction probabilities with the flux of incoming mCPs discussed in section~\ref{sec:prod}.  

As a benchmark, let us consider a millicharged particle of charge $3\times10^{-3}$, which evades existing constraints for masses above $\sim 100$~MeV. As we have seen in equation~(\ref{eq:lambda}), the mean free path for such a particle to interact above the threshold is of order ten kilometers. The probabilities for one and two hits within the ArgoNeuT volume are $\sim 10^{-4}$ and $10^{-8}$ respectively. Though both probabilities are low, during its full run, of order $ 10^9$ mCPs of mass 100 MeV would have passed through the volume of ArgoNeuT. The expectation is, therefore, $10^5$ single hit events and of order ten double-hit events. As we described above, the double-hit events will consist of two hits that are in line with the NuMI target. In the next subsection, we will show that the number of background events that meet this criterion is small and that using two hit events places a stronger limit on millicharged particles.
In our simulation of the signals, we include both the kinematic distributions of the signal energies as well as detection efficiencies extrapolated from Ref.~\cite{Acciarri:2018myr}. The efficiency used in this study starting with 0\% at 300 keV detection threshold and reaches 100\% at 800 keV with a linear increase.

\section{A Millicharge Particle search with ArgoNeuT}\label{sec:ArgoNeuT}

%The ArgoNeuT detector is located 975 meters downstream of the target in the NuMI beamline at Fermilab, with a detector geometry of,
%\beq
%L\times W\times H = 0.9~\meter\times0.4~\meter\times0.47~\meter.
%\eeq

To search for mCPs in ArgoNeuT, an estimate of the background is needed for single-hit and double-hit events. 
We will be making use of the excellent spatial resolution LAr detectors to reject double-hit events that do not point back to the target.
The ArgoNeuT detector's readout system~\cite{Anderson:2012vc} comprises of two wire planes on the $\hat y$-$\hat z$ side of the detector (with $\hat z$ roughly along the beam direction and $\hat y$ being vertical in the lab). The resolution in this plane is set by the wire spacing whereas the resolution along the horizontal $\hat x$ direction is set by the time sampling of the electronics and the electron drift velocity and is thus more precise. We will approximate the ArgoNeuT resolution in cartesian coordinates
with
\beq
%\Delta z\times \Delta x\times \Delta y = 5~\mm\times0.3~\mm\times5~\mm.
\delta y\times \delta x \times \delta z = 5.6~\mm\times0.3~\mm\times3.2~\mm.
%\Delta L\times \Delta W\times \Delta H = 4.9~\mm\times0.31~\mm\times4.9~\mm.
\eeq
where the difference in the resolution between the $z$ and the $y$ directions is due to the 60$^\circ$ angle of the wires with respect to $\hat z$ in the $\hat y$-$\hat z$ plane~\cite{Anderson:2012vc}.
With this precise resolution, ArgoNeuT an effectively three-dimensional pixelated detector with approximately 
$80\times1300\times270\simeq 3\times10^7$ pixels which can be used in our analysis. 
%As we shall see, this spatial resolution is particularly useful for background rejection in the search for millicharged particles in multi-hit events.

The ArgoNeuT experiment has taken around 3.26 million frames, each capturing roughly $4\times 10^{13}$ POT~\cite{Acciarri:2018myr}. 
The~vast majority of these frames do not contain a charged or neutral current neutrino event and are thus dubbed ``empty frames''. Such empty frames were used by ArgoNeuT in~\cite{Acciarri:2018myr} as a control sample for a data-driven background estimate in their recent analysis of de-excitation photons that accompany neutrino interactions. This dataset can also be used as the signal region for an mCP search. Indeed, in~\cite{Acciarri:2018myr} it was found that empty frames contain significantly less MeV scale hits as compared to frames that contain neutrino interactions.  
In particular, of the empty frames,  88\%  have zero hits with the remaining frames containing one or, in rare cases, more hits. Here a hit corresponds to an energy deposition above the detector threshold around MeV. 
For our background estimates we will thus assume an average number of hits of $p_\mathrm{hit}\simeq 0.128$ per frame\footnote{Assuming the background hits are independent, the average number of hit follows a Poisson distribution. Given that 88\% of empty frames has zero hits, $p_\mathrm{hit}=-\log(0.88)\simeq 0.128$. The probability for these empty frames to have one, two or three hits are hence, 11.2\%, 0.7\%, 0.03\%, which can be used to validate if the background assumption in a calibration process.}. Since this number is smaller than unity, it can also be interpreted approximately as the probability for a hit per frame per ArgoNeuT-sized volume. With this assumption the fraction of $n$-hit frames is~$(p_\mathrm{hit}^n/n!)$ following from Poisson statistics.

With these assumptions the number of single-hit events in the ArgoNeuT dataset is
\begin{equation}
    \label{eq:1-hit}
    N_\mathrm{1\,hit} = N_\mathrm{frames}\times \sum_{n} n\times \exp(-p_\mathrm{hit})\frac{p_\mathrm{hit}^n} {n!} = 4.2\times 10^5\,.
\end{equation}
The number of double-hit events will be a factor of $p_\mathrm{hit}/2$ smaller. However most double hit events can be rejected due to mis alignment with the target. 
Considering the first hit in a double hit event, only a small fraction of the detector volume, of order $(\delta x\, \delta y/\Delta x \Delta y)$,  will be appropriately aligned.
Assuming both hits are randomly distributed in the detector volume the number of the number of double hit events that are aligned with the target will approximately
% \begin{equation}
% \label{eq:2-hit}
% N_\mathrm{2\,hit} \sim \left(  \frac{\delta x} {\Delta x} \frac {\delta y}{\Delta y} \right)
% \frac{p_\mathrm{hit}^2}{2} 
% N_\mathrm{frames}  \sim 0.2\,.
% \end{equation}
\begin{equation}
    \label{eq:2-hit}
    N_\mathrm{2\,hit} = N_\mathrm{frames}\times \sum_{n} \binom{n}{2}\times \exp(-p_\mathrm{hit})\frac{p_\mathrm{hit}^n} {n!} = 2.7\times 10^4\,,
\end{equation}
amongst which the number of doublet background aligned to the target are,
\beq
N_\mathrm{2\,hit}^\mathrm{aligned}=N_\mathrm{2\,hit}\times \left(  \frac{\delta x} {\Delta x} \frac {\delta y}{\Delta y} \right)=0.24.
\eeq
To summarize, in going from single to double hit events the signal rate for the $\epsilon=3\times10^{-3}$ benchmark has decreased by a factor of $L/\lambda \sim 10^{-4}$ while the background rate has dropped by $\sim 10^{-6}$, mostly due to the high spatial resolution in liquid argon.\footnote{We note that the induced uncertainty from spatial resolution of $\delta x$ and $\delta y$ when projected back to the target location also depends on the doublet seperation in the $z$ direction. The doublets that are adjacent in the $z$ direction, their uncertainties will be larger. After convoluting with the $z$ direction distributions of the backgrounds, the expected number of background events becomes around 1.8.}

To estimate the sensitivity of this search strategy we plot in Figure~\ref{fig:exclusion}  the expected limit ArgoNeuT can place on the mCP parameter space. We consider both single and double-hit limits. For the single-hit limit (blue), we require that the number of signal events does not exceed the number of single-hit events seen by ArgoNeuT, shown in Equation~(\ref{eq:1-hit}). For the double-hit limit, since the background is expected to be low, we draw a contour in the charge-mass plane that corresponds to 3.5 double hits events in the ArgoNeuT run, corresponding to a 95\% CL limit with Poisson statistics.
\begin{figure}[tbp]
    \begin{center}
        \includegraphics[scale=0.6,clip]{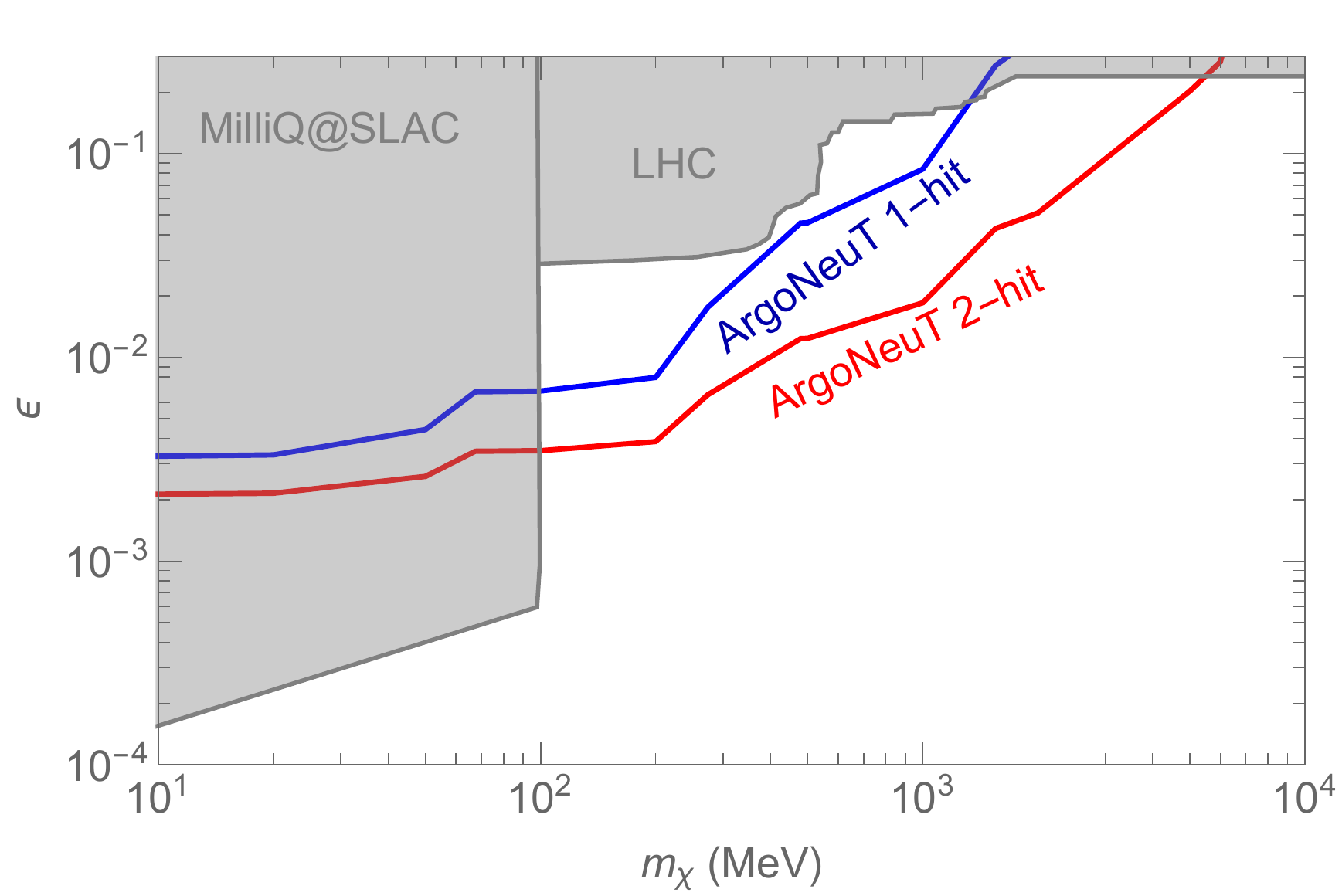}
        \caption{The physics reach in the $m_\chi$-$\epsilon$ plane for  millicharged particles achievable by the ArgoNeuT experiment with existing data with $10^{20}$ POT. The reach of a single-hit analysis is shown in blue and that of a double-hit analysis, requiring that the two hits line up with the target, is shown in red. Existing limits from other experiments, including SLAC MilliQ~\cite{Prinz:1998ua} and collider experiments~\cite{Vogel:2013raa,Essig:2013lka,CMS:2012xi,Jaeckel:2012yz},  are shown in grey. 
        }
        \label{fig:exclusion}
    \end{center}
\end{figure}
Also shown in this figure are the current limits on mCPs from SLAC MilliQ experiment and the LHC.
We see significant improvement in the coverage of mCPs in the $\epsilon$ around $10^{-2}$ for mCP mass of above 100~MeV from both searches. Especially, as we have argued the definite advantage from the double hit search strategy, the physics reach shown in red improves the result of a possible single hit search significantly. This striking physics from the relatively small scale detector of ArgoNeuT with its calibration data shows how a neutrino detector can be simultaneously viewed as a new particle physics detector, owing to its clean environment and high intensity of the beam.

In a more careful analysis, the estimate of equation~(\ref{eq:2-hit}) for the number of aligned two-hit events should be refined. For example, one may wish to allow for a looser cut on the angle between the two hits.
As we have seen in equation~(\ref{eq:scattering-angle}) a mis-alignment of up to $10^{-3}$ can be expected due to the recoil of the mCP in the first hit. In addition, matter effects described in equation~(\ref{eq:matter}) lead to a deviation in the arrival angle that is a few times larger. One can imagine a case in which the suppression factor for aligned double hits is dominated not by the finite spatial resolution but rather by the angular cut on the angle between the two hits $\theta_\mathrm{cut}$. To account for this  in equation~(\ref{eq:2-hit}) we would substitute
\beq
\label{eq:2-hit2}
\left(  \frac{\delta x} {\Delta x} \frac {\delta y}{\Delta y} \right) \to \mathrm{max}\left\{  \left( \frac{\delta x} {\Delta x} \frac {\delta y}{\Delta y} \right) \,  , 
\left(\frac{\theta_\mathrm{cut}^2}{2\pi}  \right)       \right\}
\eeq
Even if one takes a generous angular cut of $\theta_\mathrm{cut}=10^{-2}$, the estimate above only goes up by a factor of a few. 
Finally, we comment that the uncertainties in the angular measurement require careful consideration and, for example, angle measurements are expected to be more precise in an event where the distance between the hits is considerable. As a result, one may wish to discard events with non-isolated double hits, those where the distance between the hits is small. This requirement may result in a 10-20\% decrease in signal efficiency. 

%
%In Figure~\ref{fig:exclusion} we show the coverage of the single hit and double hit search strategy for mCPs with the {\it existing} ArgoNeutT data in blue and red lines, respectively. Also shown in this figure is the current limits on mCPs from SLAC MilliQ experiment and the LHC. We show the significant improvement in the coverage of mCPs in the $\epsilon$ around $10^{-2}$ for mCP mass of above 100~MeV from both searches. Especially, as we have argued the definite advantage from the double hit search strategy, the physics reach shown in red improves the result of a possible single hit search significantly. This striking physics from the relatively small scale detector of ArgoNeuT with its calibration data shows how a neutrino detector can be simultaneously viewed as a new particle physics detector, owing to its clean environment and high intensity of the beam.

\section{Outlook for DUNE ND and other Liquid Argon detectors}\label{sec:DUNE}

We have seen that even a small LAr detector can place interesting new limits on mCPs. It is interesting to consider the opportunities and challenges in performing a similar search in current and future LAr detectors that are not only larger but will also be in more intense beams. For this, we will consider the DUNE ND. However many of the consideration discussed here may apply to current and near-future SBN experiments such as microBooNE, SBND, and ICARUS which may consider mCP signals from the booser~\cite{Magill:2018tbb} but also the NuMI beamline~\cite{Coloma:2015pih, deGouvea:2018cfv}.
To estimate the reach at the DUNE ND, we will scale up the signal and background estimates for ArgoNeuT described in the previous sections.

%As a first pass, we can scale up the entire background and signal rates from ArgoNeuT to DUNE. 
%We assume DUNE LAr near detector (ND) to be a $5\times 3\times 5$~meter$^3$ detector placed 576 meters from the target with a total exposure of $3\times10^{22}$ POT over its full run. We will comment on the utility of other planned detector components near the end of this section.

%Starting with the signal, 
The number of millicharged particles passing through the DUNE ND in its entire run is shown in the right panel of  Figure~\ref{fig:prod}. 
From this, it is straightforward to calculate signal rates for single-, double-, and even triple-hit events using the formulae in section~\ref{sec:probabilities}. The ratios of single to double and that of double to triple hit events, $\sim L/\lambda$, will obviously be not as small as those for ArgoNeut due to the larger dimension of the detector. 
Below we will show expectations for single-hit and double-hit analyses, and mention triple-hit analyses in passing. The reason for this is that the double-hit backgrounds are small for most of the benchmarks we consider and going to triple-hits will not benefit the reach\footnote{Should the background uncertainty be systematically dominated, the triplet is still advantageous for heavy mCPs.}.
As a benchmark, consider a mCP with $m_\chi=200$~MeV and a charge of $10^{-3}$. For this point, of order $10^{12}$ mCP pass through the DUNE ND and the mean free path is of order 100~km, yielding roughly $10^8$ single-hit events,  $10^4$ double hit events.\footnote{Note that if one selects rare mCPs hits with higher energy transfer, the single hit induces an energetic electron with a track. It might resemble the charged current of the neutrino interaction and hence buried under neutrino events. Still, spectral information of the mCPs hits shall be of interests for improving single hit analysis result if one can distinguish.}

Extracting a potential reach for mCPs in DUNE ND requires an estimate of the backgrounds in this future facility. 
For this, we will use the background rates in ArgoNeuT as a reference and scale it to the DUNE ND. %Since the source of the ArgoNeuT empty frame background is not yet known, \ZL{The current ArgoNeuT background in the empty frames are mainly from}
The ArgoNeuT background is composited from different sources, 
ambient gamma-ray activity, intrinsic Ar$^{39}$ activity, photons produced by entering neutrons from neutrino interactions occurring upstream of the detector, and electronics noise~\cite{Acciarri:2018myr}.
%radiation from LAr, neutrino induced background, neutron background as well as the instrumental background.
Since it would be hard to model at this point which background will dominant at DUNE ND, we will consider two possible benchmarks for how these backgrounds scale to DUNE ND:
\begin{itemize}
\item \emph{Volume scaling}: we assume that the background rate per frame scales with the detector volume. With a detector volume that is 350 times larger than ArgoNeuT, with this scaling we expect about 45 soft hits in every frame. Assuming about $10^8$ frames, the total number of hits in the DUNE ND with this assumption is $4.5\times 10^{9}$. This scaling behavior may be expected if the backgrounds are detector related.
\item \emph{Volume and beam-intensity scaling}: If the background observed by ArgoNeuT are dominantly related to the beam,  one may expect that their rate scales with the detector volume and the beam intensity. Assuming a beam at the DUNE ND that is roughly forty times as intense than at ArgoNeuT (a factor of ten for beam power and a factor of four from the distance to target), in this scaling, we get about 1800 hits per frame and $1.8\times 10^{11}$ total hits.
\end{itemize}
In the figure below we will show sensitivity \emph{bands} in the mCP parameter space that span these two benchmarks for the background rates at DUNE ND.

We wish to note that these two options for background scaling are rough guesses that may be off in either direction. For example,
here we only scaled up the background rate observed by ArgoNeuT in empty frames, those without a neutrino event. In the DUNE ND, inherently all frames will contain neutrino events. In~\cite{Acciarri:2018myr} it was shown that considering events with a neutrino charged current interaction without pions or showers lead to a higher hit multiplicity, about one hit per frame, due to de-excitation photons and neutrons. 
Here we will not attempt to include these contributions and will assume that tracks and showers identified from a neutrino event can be subtracted.  
It is to be expected that the hit multiplicity will be higher given the longer drift time for DUNE ND in comparison with ArgoNeuT due to a larger detector and larger beam intensity. 
On the other hand, it is interesting to recall that ArgoNeuT is not much bigger than the absorption length of neutrons in LAr. If the source of the ArgoNeuT background is dominantly neutrons, the central region of the DUNE ND may be self-shielded. Enhanced sensitivity may be possible by choosing a cleaner fiducial volume in the center or downstream side of the detector. Besides, if electronic noise plays a role in the ArgoNeuT background, the more advanced electronics of DUNE ND may lead to lower rates. As data best guide these considerations, we do not attempt to derive a more refined estimate and proceed with the two benchmarks above.

With these assumptions for signal and background rates, several analyses may be considered. 
We will consider the single hit analysis and three different double hit analyses which will be described below and yield improved sensitivity compared to single hits:
\paragraph*{\underline{Single-hit analysis}:} This analysis is similar to that in ArgoNeuT. The blue band in the top panel of Figure~\ref{fig:exclusionDUNE} shows the sensitivity of requiring that the number of single hit events is smaller than the total assumed background rate. In fact, under the volume$\times$POT background scaling assumption, both signal and backgrounds scale in the same way and the DUNE ND single-hit sensitivity is essentially identical to that of ArgoNeuT. It is possible that single hit backgrounds will be lower than our benchmarks, say, using a fiducial volume, and any further background reduction of the background can be carried out, and the limits will improve by the one-fourth power of the background reduction.
Even with the more favorable background assumption, we see that ArgoNeuT's double hit sensitivity will likely exceed that of DUNE ND single-hit analysis.

%We observe that in the conservative background of scaling with volume and beam-intensity the single hit limit is worse than the ArgoNeuT due to the larger background over signal ratio. The lower part of the blue band shows the result of single hit search if the background scales with detector volume only, which is a factor of 40 smaller backgrounds comparing to the conservative assumption.  

\paragraph*{\underline{Double-hit analyses}:}
For the double-hit signal, we will consider three possible analysis strategies. In all cases, we will reject backgrounds using the requirement of alignment of the two hits with the target, as in equations~(\ref{eq:2-hit}) and~(\ref{eq:2-hit2}) which leads to a factor of order $10^{-7}$ reduction in backgrounds with the angular cut, $\theta_\mathrm{cut}\sim 10^{-3}$. We note however that given a frame with $n$ hits, number of pairs of hits is of order $n^2/2$. For the two background benchmarks above this corresponds to around $4.6\times 10^3$ to $7.1\times 10^{6}$ pairs per beam spill, and $10^8$ times that over the full dataset. 
In the bottom panel of Figure~\ref{fig:exclusionDUNE} we show three different sensitivity bands corresponding to different approaches to mitigating this high occupancy background:
\begin{itemize}
\item \emph{Systematic limited}: 
In the most conservative approach, which we show for comparison's sake, we assume the background is systematically limited and draw the sensitivity curve where the number of signal double-hit events $S$ is equal to the background prediction $B$, excluding $S\geq B+2\sqrt{B}$. For our background benchmarks, this corresponds to $1.4\times 10^4$ to $2.2\times 10^7$ double-hits that are in line with the target. This result is shown as the dashed red band in Figure~\ref{fig:exclusionDUNE}. We expect that the double-hit background will be under control and hence our statistically limited analysis is thus more realistic.

\item \emph{Statistics limited}: 
Even though the source of backgrounds may be unknown, the use of double hits allows for additional handles to predict the background aligned with the target. In particular, if the background hits are uncorrelated, one may predict the rate of double hits aligned with the target by extrapolating from the adjacent angular bins which will be equally populated. The result is that the limit on new physics becomes limited by statistical rather than systematic uncertainty and the figure of merit for setting limits on mCP's is $S/\sqrt{B}$, rather than $S/B$, excluding $S\geq 2\sqrt{B}$.
This additional square-root in the number of events limited by the data counteracts the occupancy-squared enhancement of double-hit backgrounds. The corresponding sensitivity is shown in the red dash-dotted band in Figure~\ref{fig:exclusionDUNE}.
This strategy can also be used for the ArgoNeuT analysis, but since the double-hit background is already low, it yields similar results. 
\item \emph{Timing reduced occupancy}: 
The DUNE ND is planned to be segmented into around 20 modules, each of which will have a separate light collection system. The light collection allows for excellent timing resolution, of order nanosecond~\cite{DUNEND}. %$XXX$%\ZL{Not sure which reference or number to put here.}.
%\ZL{not sure if we can quote some numbers here. It should be easy to achieve 100 reduction. Derived from the x direction resolution, the timing precision should be ten picoseconds. The frame time is of order millisecond, and the spill separation is of order a second if I remember correctly.}. 
If every soft hit can be associated with a particular time, it is possible to reduce backgrounds further by requiring the two hits to be nearly simultaneous. This method effectively amounts to increasing the effective number of frames into which the events are distributed, and thus the corresponding occupancy is lower. In the solid red band in Figure~\ref{fig:exclusionDUNE} we show the sensitivity assuming the effective number of frames is a factor of 100 larger by associated the precision timing information of the hits. Here we also assume the limit is set by statistical uncertainties using angular sidebands as in the previous case. 
\end{itemize}

\begin{figure}[t]
    \begin{center}
        \includegraphics[scale=0.5,clip]{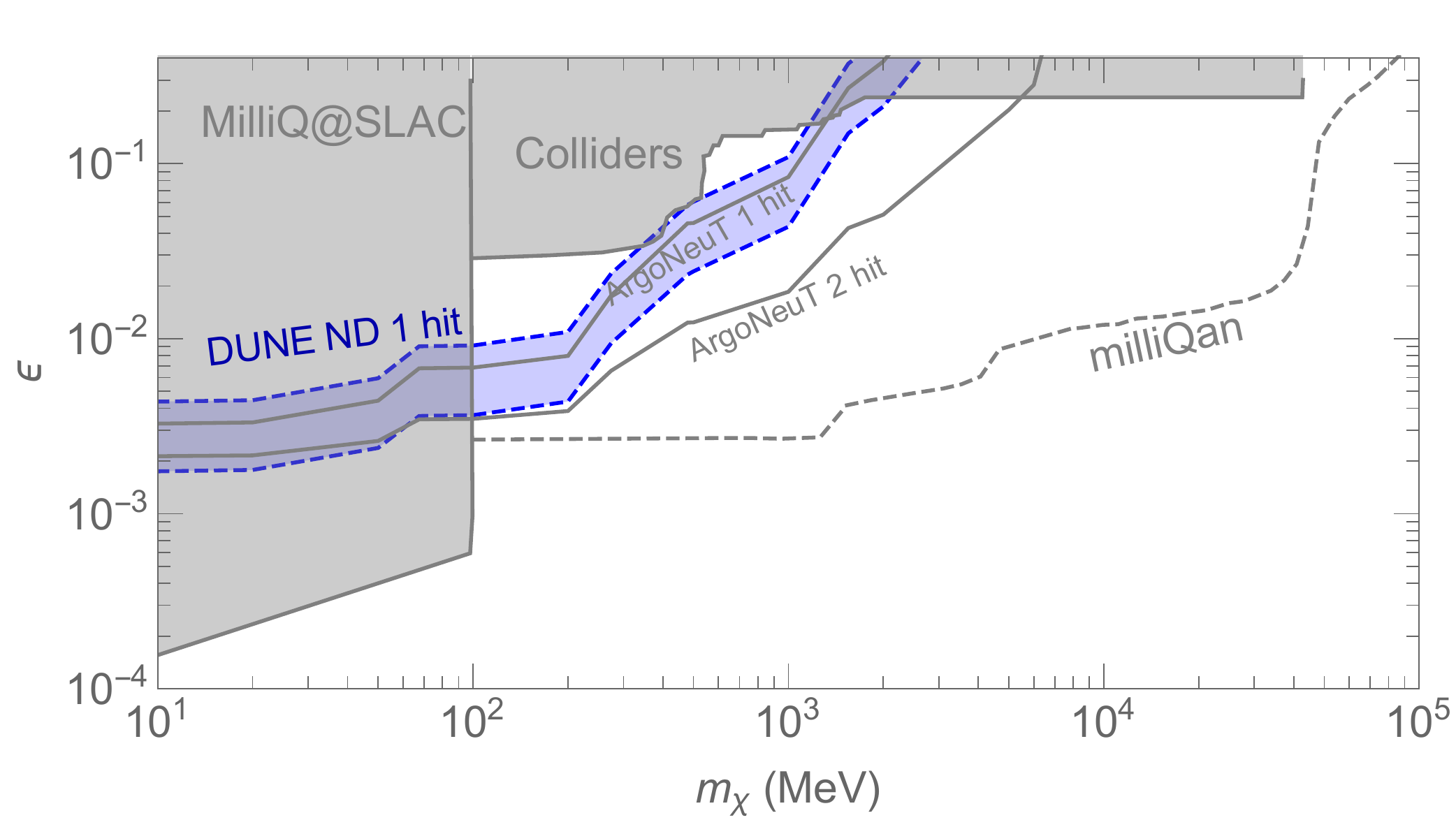}
        \includegraphics[scale=0.5,clip]{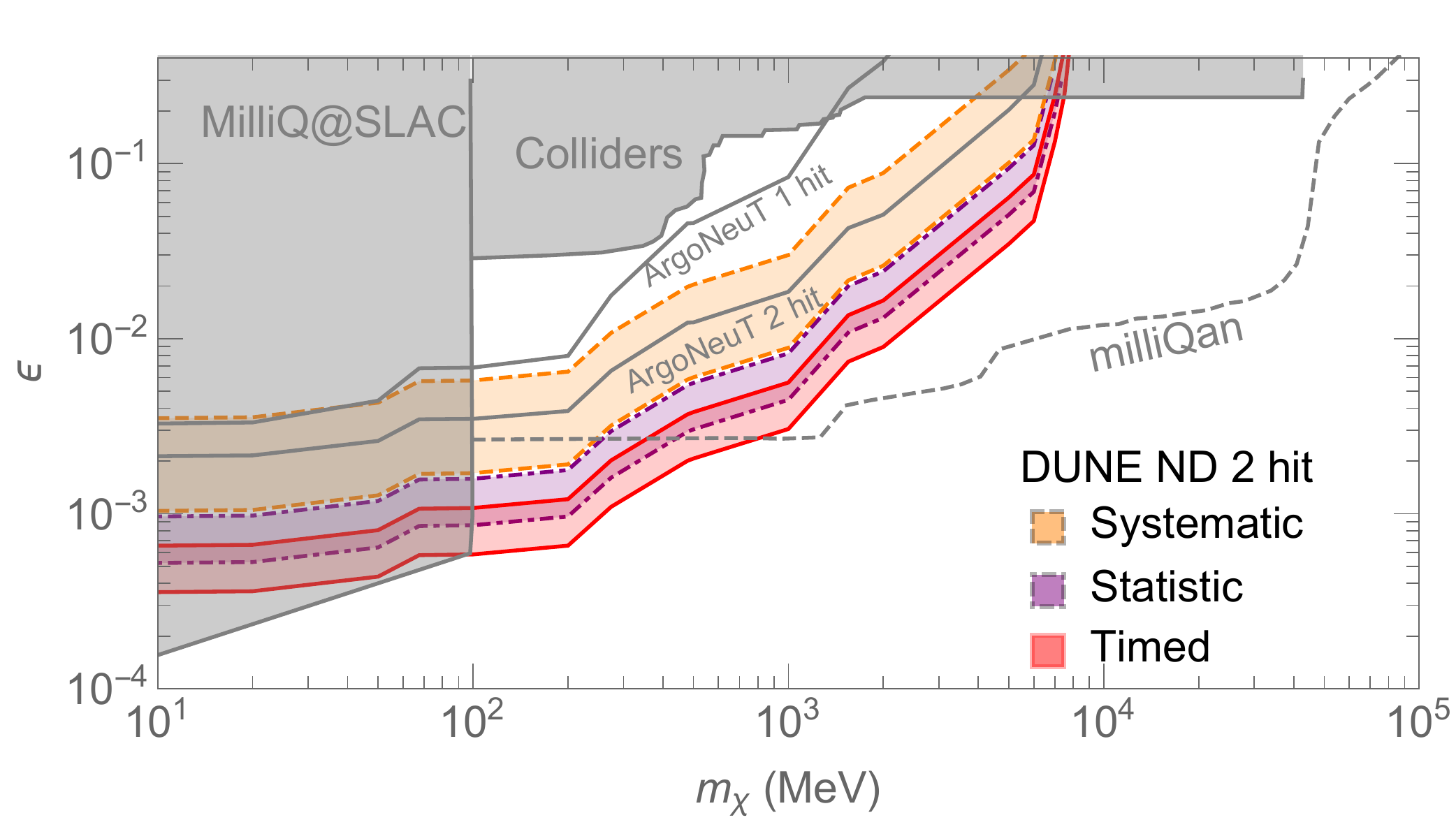}
        \caption{The reach in the $m_\chi$-$\epsilon$ plane for millicharged particles for various analyses in the DUNE ND with our projection for $3\times 10^{22}$ POT. For every analysis a band is shown which spans our two benchmark assumptions for the backgrounds: scaling only with detector volume of scaling (low BG) also with beam intensity (high BG). \emph{Top:} the sensitivity of a systematically limited single-hit analysis is shown in blue. \emph{Bottom:} The sensitivity of double-hit analyses, requiring the two hits align with the target. A systematically limited search is shown in orange. A statistically limited search (making use of the angular distribution) is shown in purple. A statistically limited search with the occupancy reduced by light collection and timing is shown in red. Existing limits from other experiments, including SLAC MilliQ~\cite{Prinz:1998ua} and collider experiments~\cite{Vogel:2013raa,Essig:2013lka,CMS:2012xi,Jaeckel:2012yz}, are shown in grey for both panels. The projections of 95\% C.L. exclusion limit from the proposed milliQan experiment at the LHC with 3000 $\fbi$ of integrated luminosity is shown in gray dashed curve~\cite{Ball:2016zrp}.
        \footnote{The milliQan experimental results might be further improved by the inclusion of the QCD production of the millicharged particles, which are the dominant production modes for low mCP masses.}
        }
        \label{fig:exclusionDUNE}
    \end{center}
\end{figure}
The inputs that go into our sensitivity bands are listed in Table~\ref{tab:DUNEbkg} for the ArgoNeuT reference and the DUNE ND background benchmarks, as well as for the various analyses. 
To summarize the prospects for DUNE ND, the double-hit signal has significant potential to go beyond the expected ArgoNeuT limit. 
The double hit background can be reduced and modeled using a data-driven method by studying the sideband, improving the results significantly, as shown in the purple band of Figure~\ref{fig:exclusionDUNE}.
Furthermore, as discussed above, the inclusion of the timing information can further improve the results as shown in the red band. Up the lower edge of the red band, the timed double hit background is of order 10 events, almost reaching a background-free search. 
We conclude that larger LAr detectors, particularly the DUNE ND, have a bright prospect of further searching for mCPs beyond the ArgoNeuT search.  

\begin{table}[htbp]
    \centering
    \caption{Background considerations. The DUNE ND background is scaled from ArgoNeuT background with two options, Scaling A (scaled with the number of frames and detector volume) and Scaling B (scaled with POT, detector volume and inverse square of distance) as specified in the text. The number of background singlet, doublet and triplet events with and without alignment with the beampipe requirement are listed in the later part of this table. Three DUNE ND background reduction options are considered, including no reduction, reduced to statistical uncertainty only, and further reduced by using timing information, labeled by ``Systematic'', ``Statistic'' and ``Timed'', respectively.
    \footnote{
        The average number of clusters(hits) per frame (without the neutrino events) are 0.13, 45, 1800 for ArgoNeuT, DUNE (Scaling A), and DUNE (Scaling B), respectively. The ``Timed'' DUNE case have a 100 times smaller average number of clusters (hits) per frame.
        %\ZL{There is a beautiful convergence of the results through two independent ways of calculating the backgrounds. I am pleasantly satisfied with the pixelated treatment and calculation on the backgrounds done here.}
        }
    %\footnote{The geometry of DUNE near detector is set as $4m \times 3{\rm m}\times 5m$ with resolution the same as ArgoNeuT.}
    }
    \bgroup
         \def\arraystretch{1.5}%  1 is the default, change whatever you need
      \begin{tabular}{|c|c|c|c|c|c|c|l|l|l|l|l}
  \cline{1-11}    \multirow{2}[0]{*}{} & \multicolumn{1}{c|}{\multirow{3}[0]{*}{Bkg Scaling}} & \multicolumn{1}{c|}{\multirow{3}[0]{*}{Bkg reduction}} & \multicolumn{3}{c|}{\# frames with} & \multicolumn{5}{c|}{\# Background events} &  \\
  \cline{4-11}      &   &   & \multirow{2}[0]{*}{$\geq0$ hit} & \multirow{2}[0]{*}{$\geq1$ hits} & \multirow{2}[0]{*}{$\geq2$ hits} & \multirow{2}[0]{*}{Singlets} & \multirow{2}[0]{*}{Doublets} & Aligned  & \multirow{2}[0]{*}{Triplets} & Aligned &  \\
   &   &   &  &  &  &  &  & doublets  &  & triplets &  \\
  \cline{1-11}    ArgoNeuT & Reference & Systematic & $3.3\times 10^6$ & $3.9\times 10^5$ & $2.4\times 10^4$ & $4.2\times 10^5$ & $2.7\times 10^4$ & 0.24 & $1.1\times 10^3$ & $9.1\times 10^{-8}$  &  \\
  \cline{1-11} 
    \multirow{6}[0]{*}{ DUNE ND} & \multirow{3}[3]{*}{Volume} & Systematic & \multicolumn{3}{c|}{\multirow{2}[0]{*}{$1\times 10^8$}} & \multirow{3}[0]{*}{$4.5\times 10^9$} & \multirow{2}[0]{*}{$1.0\times 10^{11}$} & $1.4\times 10^4$ & \multirow{2}[0]{*}{$1.6\times 10^{12}$} & 0.030 &  \\
  \cline{3-3}\cline{9-9}\cline{11-11}      &   & Statistic & \multicolumn{3}{c|}{} &   &   & $\sqrt{1.4\times 10^4}$ &   & 0.030 &  \\
  \cline{3-6}\cline{8-11}      &   & Timed & $1\times 10^{10}$ & $3.6\times 10^{9}$ & $7.6\times 10^{8}$ &   & $1.0\times 10^9$ & $\sqrt{1.4\times 10^2}$ & $1.6\times 10^{8}$ & $3.0\times 10^{-6}$ &  \\
  \cline{2-11}   
     & \multirow{3}[3]{*}{Vol.$\times$Int.} & Systematic & \multicolumn{3}{c|}{\multirow{2}[0]{*}{$1\times 10^8$}} & \multirow{3}[0]{*}{$1.8\times 10^{11}$} & \multirow{2}[0]{*}{$1.6\times 10^{14}$} & $2.2\times 10^7$ & \multirow{2}[0]{*}{$9.3\times 10^{16}$} & $1.8\times 10^3$ &  \\
  \cline{3-3}\cline{9-9}\cline{11-11}      &   & Statistic & \multicolumn{3}{c|}{} &   &   & $\sqrt{2.2\times 10^7}$ &   & $\sqrt{1.8\times 10^3}$ &  \\
  \cline{3-6}\cline{8-11}      &   & Timed & \multicolumn{3}{c|}{$1\times 10^{10}$} &   & $1.6\times 10^{12}$ & $\sqrt{2.2\times 10^5}$ & $9.3\times 10^{12}$ & 0.18 &  \\
  \cline{1-11}    \end{tabular}%
  \egroup
    \label{tab:DUNEbkg}%
\end{table}%

\subsubsection{Other near detectors and dedicated strategies}
Though the potential sensitivity shown above is promising, so far we have exploited the capabilities demonstrated in LAr TPCs for neutrino physics. It is interesting to consider other advanced detector technologies as well as dedicated searches in LAr. 
\begin{itemize} 
\item \emph{Argon Gas TPC:}
To mitigate the high occupancy in the DUNE ND  and to study events in greater detail, a high-pressure gaseous argon TPC detector is proposed. 
Naively, argon gas, with its lower density, will produce less signal event because the mean free path for an mCP is of order $(\rho\sigma)^{-1}$, where $\rho$ is the detector material density.
However, precisely because of the lower density, charged particles travel in gas farther than they do in liquid argon~\cite{Tanabashi:2018oca}. As a result, the effective energy threshold will be much lower, as low as 10~keV. As we showed in Equations~(\ref{eq:sigma}) or~(\ref{eq:lambda}), the scattering cross section is inversely proportional to the detection threshold. The total signal event rate in a gas TPC is thus expected to be parametrically similar to that in a liquid detector. The backgrounds in a gas detector may, however, be significantly lower, potentially leading to enhanced sensitivity. Further improvements may be achieved by combining hits in the LAr and the gas TPC that align with the distant target, as well as making use of the planned electromagnetic calorimeter. 

\item \emph{Off axis detectors:}
The dominant source of soft hits in near detectors is likely to be beam-related. Since the charged pions are collimated in the magnetic horns, the backgrounds may be highly peaked in the forward region. In~\cite{Coloma:2015pih} and~\cite{deGouvea:2018cfv} it is shown that dark sector signals which are produced by the decay of neutral particles are produced in a much wider beam than the neutrino beam and that and thus the signal to background ratio is higher in off-axis detectors. MicroBooNE and ICARUS, two LAr detectors, are located about~$6^\circ$ off-axis, an ideal angle for this purpose. The angular distribution of mCPs is similar to that of secluded dark matter and is shown in~\cite{deGouvea:2018cfv}. Within the LBNF beam, the proposed DUNE-PRISM detector~\cite{PRISM} will also be able to cover off-axis angles.

\item \emph{Faint tracks - a dedicated millicharged reconstruction:} 
The standard track reconstruction in LAr TPCs is geared towards charged particles. Initially, hits are identified by finding individual wires in which charge is detected with a certain significance above the noise. Hits are then collected into clusters which are then further grouped as tracks. Crucially for our discussion, individual hits are identified independently from one another. As a result, the threshold for each charge deposition needs to be high enough to exceed the noise pedestal independently of any other. However, if one searches for a correlated excess of deposited charge along a line, the noise along the line can be integrated down, and a lower local threshold can be set. Naively the threshold may be lowered by $\sqrt{N}$ where $N$ is the number of ``pixels'' along a track that traverses the detector, which can be of order a thousand.

\item \emph{A dedicated detector:}  One may consider installing a dedicated detector to search for mCPs in a neutrino beam. Indeed, the successful SLAC milliQ experiment~\cite{Prinz:1998ua} is an example of a dedicated experiment in a fixed target setup. The milliQan detector concept~\cite{Ball:2016zrp} uses plastic scintillator bars with good timing resolution to look for triple-coincident photoelectron production. This proposed detector is expected to have good efficiency for mCP detection down to charges of a few$\times 10^{-3}$. The reach of such a detector in the Fermilab beams was recently studied in~\cite{Kelly:2018brz}. It is interesting to note that the scintillation technology and LAr are complementary, one relying on good timing resolution while the other employing good spatial resolution. A combination of the two detector technologies may be beneficial. Indeed one of the components of the planned DUNE ND is a 3D scintillator tracker and its utility in new particle searches should be studied.  
\end{itemize}

\section{Conclusions}\label{sec:outlook}

In this work, we explored the potential of LAr TPC detectors to search for mCPs. We found that the recently demonstrated capability of ArgoNeuT to detect MeV scale energy deposits in LAr~\cite{Acciarri:2018myr} enables a very sensitive search. We have also shown that backgrounds can be reduced using double-hit events in which the two hits point back to the distant proton target. This new proposal yields an enhanced sensitivity in the mCP parameter space. We find that the ArgoNeuT, a small pioneering LAr detector, can probe regions of parameter space that are unexplored with its existing dataset. 

We have also speculated about the reach of future LAr detectors. Assuming the background rate scale with detector volume and beam intensity, the high occupancy in these detectors leads to large two-hit backgrounds, though sensitivity can still be improved using a data-driven background estimation which is statistically limited. Using light collection in LAr to improve timing resolution, or using a gaseous argon TPC may lead to a further reduction of backgrounds and improved sensitivity.

\acknowledgments{We are very grateful to Ivan Lepetic for many discussions and insights on ArgoNeuT's capability for detecting MeV recoils which enabled this work. We also thank Jen Raaf and Bryce Littlejohn for useful discussions and insights. Fermilab is operated by Fermi Research Alliance, LLC under Contract No. DE-AC02-07CH11359 with the U.S. Department of Energy. The authors thank the Aspen Center for Physics, which is
supported by National Science Foundation grant PHY-1607611. 
ZL is supported in part by the NSF under Grant No. PHY1620074 and by the Maryland Center for Fundamental Physics.
%The authors thank the Aspen Center for Physics, which is supported by National Science Foundation grant PHY-1607611.
}

\clearpage

\appendix

%\section{Supporting material}
%\ZL{Does the focusing horn has any deflection effect on the mCP flux?}
%\ZL{Comment on the spectral information available as well.}
%\ZL{For mCP hard scattering in the detector, does it look exactly like a neutrino charged current event?}

\section{Scattering angle calculation}
\label{sec:deflection}
In section~\ref{sec:angle} we estimated the angular deflection of mCPs traveling through matter. It is well known that, as opposed to energy loss, the angular deflection of charged particles moving through matter is dominated by multiple soft scatterings off of nuclei~\cite{Tanabashi:2018oca}. Since there is no correlation between one scattering and the next, the total deflection can be described as a random walk. 
The differential scattering cross section of an mCP off a nucleus is similar to Equation~(\ref{eq:dsigmader}), but with the nucleus mass $m_A$ replacing the electron mass and with an additional factor of $Z^2$ to account for the large nuclear charge. The IR divergence of this formula is cut off by the atomic form factor, which we will approximate here as a sharp cutoff. However, we will see that the results are insensitive to the value of the IR cutoff. 
With every scattering in which the nucleus recoils by $E_r$ the angle of the mCP is deflected by 
\beq
\theta_\chi (E_r) \sim \frac{\sqrt{2 m_N E_r}}{E_\chi}
\eeq
The average angular deflection in a nuclear scattering is
\begin{equation}
\langle \theta_\chi \rangle = \frac{1}{\sigma (E_\mathrm{min})} \int_{E_\mathrm{min}}^{E_\mathrm{max}} \frac{\partial \sigma_\mathrm{nuc} }{\partial E_r} \theta_\chi (E_r)  dE_r     
\end{equation}
In traversing a length $L$ of material, the number of such collisions is simply the $L/\lambda$ where $\lambda$ is the mean free path derived in analogy to Equations~\ref{eq:dsigmader}-\ref{eq:meanfreepath}, with the appropriate replacements of $m_e\to  m_N$ and the additional coherent factor of $Z^2$ instead of the incoherent $Z$. 
The total deflection  due to $N_\mathrm{col}$ collisions is 
\begin{equation}
\Delta \theta_\chi \sim \langle\theta_\chi\rangle \sqrt{N_\mathrm{col}}
\end{equation}
Evaluating this expression numerically for a range of nuclei gives Equation~(\ref{eq:matter}). It is interesting to note that this estimate is insensitive to the IR cutoff since the square root of the number of collisions scales as $E_\mathrm{min}^{-1/2}$ while the average deflection scales as $E_\mathrm{min}^{1/2}$. The total deflection is also quite insensitive to the nuclear species.

\bibliographystyle{utphys}%{plainnat}
\bibliography{millicharged18}

\end{document}